\documentstyle[12pt,a4,epsfig]{article}
\def\Journal#1#2#3#4{{#1} {\bf #2}, #3 (#4)}

\def\PLB{{\em Phys. Lett.}  B}
\def\PRL{\em Phys. Rev. Lett.}

\def\PRC{{\em Phys. Rev.} C}

\def\ZPA{{\em Z. Phys.} A}

\begin{document}

\title{\vspace*{-2cm}
PION INTERFEROMETRY USING WAVEPACKETS \footnote
{Talk presented at the 2nd Catania Relativistic Ion Studies CRIS '98. To be
published in the Conference Proceedings by World Scientific.}}
\author{D. PELTE, H. MERLITZ}
\date{}
\maketitle
\small \noindent \vspace*{-1.5cm}
\begin{center}
Physikalisches Institut der Universit\"at Heidelberg,
\\Philosophenweg 12, D-69120 Heidelberg, GERMANY
\\E-mail: pelte@pel5.mpi-hd.mpg.de
\end{center} 
\normalsize
\abstract{The conditions for the HBT interferometry from stars
are compared to those encountered in heavy ion reactions. As a consequence
the formalism using wavepackets is developed. The modifications relative to
the standard formalism with plane waves are presented, and the effects of
the residual Coulomb interactions between charged pions and with the charged
source are calculated.}

\section{Introduction}
Originally the interferometric method of Hanburry Brown, Twiss (HBT) was
developed in astronomy and used to determine the angular radii of stars
~\cite{han57}. Shortly afterwards it was recognized that a similar procedure
can also be applied to pions emitted in high energy reactions to measure the
size of the pion emitting source ~\cite{gol60}. However, it is clear that the
different conditions prevailing in high-energy or heavy-ion reactions ask
for certain modifications of the original formalism. The table 1 lists the
relevant parameters encountered in the study of stars or hot nuclear
\begin{table}[h]
\caption{Parameters relevant to stars and nuclei.}
\vspace{0.2cm}
\begin{center}
\footnotesize
\begin{tabular}{|c|c|c|}
\hline
 & star (photon) & nucleus (pion) \\
\hline
source size R & $10^9$ m & $10^{-14}$ m \\
momentum p    & 2 eV/c   & 2 $\cdot 10^8$ eV/c \\
correlation area $\Delta \Omega = \pi / (pR)^2$ & 3 $\cdot 10^{-32}$ sr &
3 $\cdot 10^{-2}$ sr \\
de Broglie wavelength $\lambda = h / p$ & 6 $\cdot 10^{-7}$ m &
6 $\cdot 10^{-15}$ m \\
coherence length $\Lambda$ & $c \tau \approx 3$ m & $\lambda^2 / \Delta \lambda
\approx 10^{-14}$ m \\
\hline
\end{tabular}
\end{center}
\end{table}
matter. The correlation areas $\Delta \Omega$ are given by the solid angle
necessary to observe the speckle pattern in the emission of a large number
of identical bosons as displayed in Fig.1 of ref.~\cite{mer95}.
Their difference by 30. order of magnitudes
is compensated by the equivalent difference in the distance between the star,
respectively the nucleus and the interferometer. The large difference between
the source size R and the coherence length $\Lambda$ in case of a star ensures
that the star can be treated as a chaotic radiator. In general one finds for
stars $R \gg \Lambda \gg \lambda$ and the emitted photons may be described
by plane waves. In case of the nucleus one finds $R \approx \Lambda \approx
\lambda$, and wavepackets are the more appropriate description of the emitted
pions. 
 
\section{Wavepacket Formalism}
The width of a Gaussian wavepacket at time t=0 is given by $\sigma_0$ which
from the principles of quantum mechanics is related to the pion coherence
length: $\sigma_0 \approx \Lambda / 4 \pi \approx 1$ fm.At later times $t$ the
free, i.e. non-interacting wavepacket in coordinate space is~\cite{mer97}
\begin{equation}
\Psi({\bf r},t) = \left( 2 \pi s^2 \right)^{-\frac{3}{4}}
exp \left\{ \frac{i}{\hbar} \left( {\bf P r} - E t \right) - \frac{
\left( {\bf r} - {\bf R} \left( t \right) \right)^2}{4 s \sigma_0} \right\} ,
\end{equation}
where ${\bf P} , E$ are the centre momentum , energy of the wavepacket,
${\bf R}(t)$
is its spatial coordinate which changes with time according to ${\bf R}(t) =
{\bf R} + ({\bf P}/m) \cdot t$, and the dispersion of the width is given by
\begin{equation}
\sigma = |s| = \sigma_0 \sqrt{1 + \frac{\hbar^2}{4 m^2 \sigma_0^2} t^2} .
\end{equation}
Similarly the wavepacket may be described in momentum space, where it turns
out to be stationary:
\begin{equation}
\Phi({\bf p},t) = \left( 2 \pi \sigma_p^2 \right)^{-\frac{3}{4}}
exp \left\{ \frac{i}{\hbar} \left({\bf R} \left( {\bf p} - {\bf P} \right) +
\frac{{\bf p}^2}{2 m} t \right) - \frac{ \left( {\bf p} - {\bf P} \right)^2}
{4 \sigma_p^2} \right\}
\end{equation}
with $\sigma_p = \hbar / 2 \sigma_0$. It is obvious that the wavepacket
representation of the pions fulfills Heisenbergs uncertainty relation
\begin{equation}
\sigma_p \sigma \ge \frac{\hbar}{2} .
\end{equation}

In the following we will restrict our discussion to SIS energies, i.e. the
like-charge pion multiplicities are in central collisions of very heavy
systems $m_{\pi} \approx 20$, and multi pion effects can be neglected. We
parametrize the pion source function by $g(R,P) = \rho(R) f(P)$ with
\begin{eqnarray}
\rho(R) = \left( \pi R_s^2 \right)^{-\frac{3}{2}} exp \left\{ -\frac{R^2}
{R_s^2} \right\} \\ f(P) = \left( 2 \pi m T \right)^{-\frac{3}{2}} exp \left\{
-\frac{P^2}{2 m T} \right\}
\end{eqnarray}
and with the temperature $T$. The single pion observables in spatial and
momentum space are then easy to calculate and yield the identical forms as
in eqs.(5,6)
except that the relevant parameters have to be replaced by $\tilde{R}_s^2 =
R_s^2 + 2 \sigma_0^2$ and $T_{eff} = T + T_{qm}$ where $T_{qm} = \frac{\hbar^2}
{4 m \sigma_0^2}$. This implies that for small pion multiplicities the
localization $\sigma_0$ of the pions in the source does not change the
single pion distributions, but the relevant parameters are changed: The
effective source radius is modified by the pion localization, and the effective
temperature is modified by a quantum contribution which takes into account the
zero point energy of the pions.
\begin{figure}
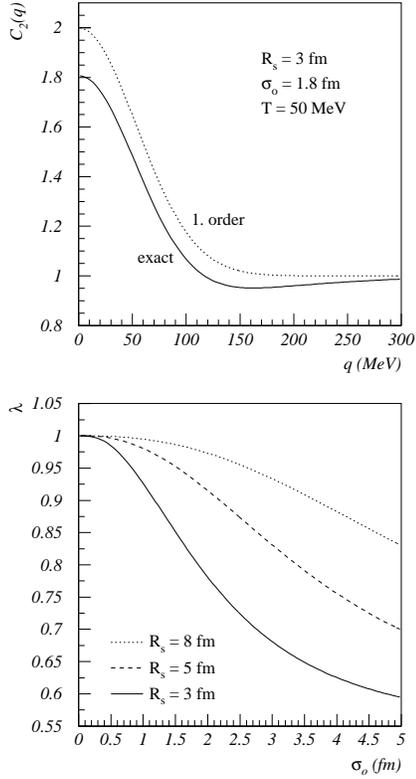

\begin{minipage}{8cm}
\epsfxsize=7cm
\epsffile[50 270 420 550]{cris98_fig1l.eps}
\end{minipage}
\begin{minipage}{8cm}
\epsfxsize=7cm
\epsffile[50 270 420 550]{cris98_fig1r.eps}
\end{minipage}
\caption{Left panel: The exact and 1. order correlation functions.
Right panel: The dependence of the chaoticity $\lambda$ on the
localization $\sigma_0$.}
\end{figure}

Similarly the two pion correlation function $C_2({\bf p}_1,{\bf p}_2) =
\frac{{\cal P}_2({\bf p}_1,{\bf p}_2)}{{\cal P}_1({\bf p}_1)
{\cal P}_1({\bf p}_1)}$ may be calculated. This calculation is more involved
than in case of the single pion observables, the exact result was published
in~\cite{mer97}. As an approximate expression one obtains
\begin{equation}
C_2({\bf p}_1,{\bf p}_2) = 1 + \lambda exp \left\{ -\frac{R_{eff}^2}{2 \hbar^2}
{\bf q}^2 \right\} \qquad , \qquad {\bf q} = {\bf p}_1 - {\bf p}_2
\end{equation}
with an effective radius parameter $R_{eff}^2 = R_s^2 + 2 \sigma_0^2 \frac{T}
{T_{eff}}$, so that $R_s < R_{eff} < \tilde{R}_s$. The chaoticity parameter
$\lambda$  is
a function of the ratios $R_s^2 / (2 \sigma_0^2)$ and $T / T_{eff}$ and can be
expanded into the series
\begin{equation}
\lambda = 1 + \sum_{k=1}^{\infty} (-)^k \left( 1 - k \frac{R_s^2}{2 \sigma_0^2}
\right)^{-\frac{3}{2}} \left( 1 - k \frac{T}{T_{eff}} \right)^{-\frac{3}{2}} .
\end{equation}
The Fig.1 displays on the left side the exact and first order ($\lambda = 1$)
results for $C_2(q)$ using specific values for $R_s$, $\sigma_0$ and $T$, and
on the right side the dependence of $\lambda$ on $\sigma_0$ is shown for 3
different values of $R_s$ and the temperature $T = 50$ MeV. From this
dependence we infer that $\lambda$ approaches 0 only when $\sigma_0$ becomes
very large and $T$ goes to zero.

The wavepacket formalism for pions introduces a new parameter $\sigma_0$ which
was called localization because it determines the probability to localize
the pion within a certain volume
which is usually assumed to be within 
the hot nuclear matter. Naively the value
of $\sigma_0$ is therefore experimentally bounded by the two limits $R_{eff} =
\sqrt{2} \sigma_0$ and $T_{eff} = T_{qm}$ which yields 0.8 fm $< \sigma_0 <$
4 fm under SIS conditions. However it is more appropriate to relate $\sigma_0$
to the coherence length $\Lambda$ of the pion, which in the case that the pion
is emitted in the decay of the $\Delta(1232)$ resonance is coupled to the
lifetime of that resonance in hot nuclear matter. Alternatively one may also
consider the mean free path of pions in hot nuclear matter as the appropriate
quantity. We have used this latter conjecture in our calculations and have
assumed $\sigma_0$ = 1.8 fm. This implies a quantum contribution of $T_{qm}$
= 21 MeV to the effective temperature of pions.
The plane wave limits are obtained in two ways: $\sigma_0 \rightarrow \infty$
(sharp momentum states) yields $C_2({\bf p}_1,{\bf p}_2) = 1 + \delta
({\bf p}_1-{\bf p}_2)$, in contradiction to experiment, $\sigma_0 \rightarrow 
0$ (sharp position states) yields an infinite zero point energy, also in
contradiction to experiment. This again demonstrates the inherent difficulties
to describe the HBT interferometry with pions in a plane wave formalism.
The wavepacket formalism was applied in other recent papers to the HBT
interferometry with pions ~\cite{mer97,wie97,wie98,zim97,cso97}.
In cases where $\sigma_0$ was specified its value was set to $\sigma_0 \approx$
0.7 fm which yields $T_{qm} \approx$ 100 MeV, a value which appears to be too
high for the SIS energy range.

\section{The Coulomb Residual Interactions}
HBT interferometric studies in relativistic heavy-ion reactions are usually
performed with charged pions. In this case the pions after emission cannot be
described by free wavepackets but experience residual interactions from the
charge of the source and from their own charges. As a result the centre motion
 of the wavepacket is modified and the wavepacket gets deformed. There are
two ways to study these effects with respect to HBT interferometry, both ways
are based on the time dependent Schr\"odinger equation: The
multiconfigurational method~\cite{mer97a}, and the molecular dynamic
method~\cite{mer97}.
\begin{figure}
\epsfxsize=16cm
\epsffile[50 300 550 530]{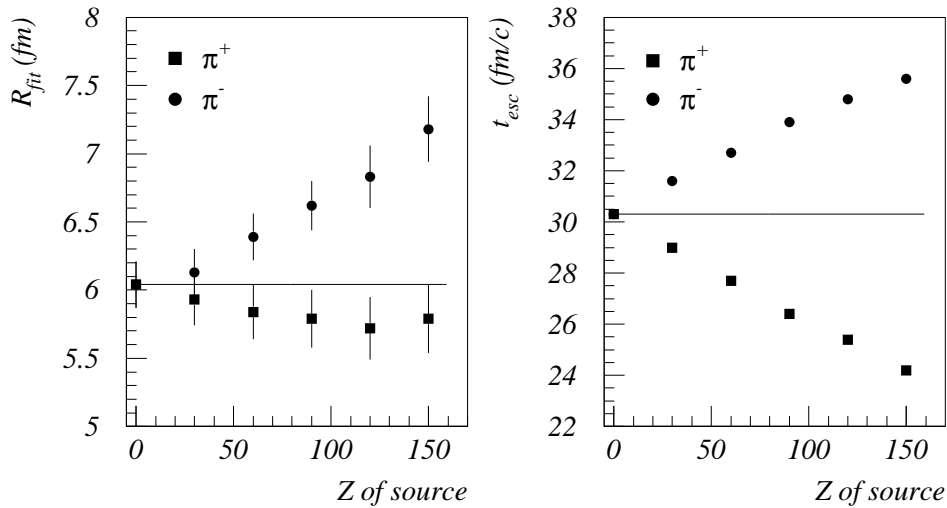}
\caption{Left panel: The dependence of the effective source radii from
($\pi^-$,$\pi^-$) and ($\pi^+$,$\pi^+$) pairs on the charge $Z$ of the source.
Right panel: The escape times $t_{esc}$ for $\pi^-$ and $\pi^+$ as functions
of $Z$.}
\end{figure}

\subsection{Coulomb Interaction with the Source}
The second method was used to study the modifications of the two-pion
correlation function when both pions interact with the source via the
Coulomb force. Depending on the charge of the source the width of the
correlation function is decreased for ($\pi^-$,$\pi^-$) but increased for
($\pi^+$,$\pi^+$) pairs, yielding a larger effective source radius from the 
former pairs, and a smaller radius from the latter pairs. The situation is
depicted in the left panel of Fig.2 for a source with radius $R_s$ = 5.5 fm and
temperature $T$ = 50 MeV. Surprisingly the increase respectively decrease are
not of same size, but stronger for the ($\pi^-$,$\pi^-$) pairs. The reason
for this asymmetry is found in the right panel of Fig.2 where the average
time is shown which a charged pion spends within the escape radius $R_{esc}$
= 20 fm of the source. In case of the ($\pi^+$,$\pi^+$) pairs the
correlation signal develops to a large fraction after the $\pi^+$ have left
the interaction region, and it is therefore only little influenced by this
interaction. In case of the ($\pi^-$,$\pi^-$) pairs the times the $\pi^-$
spend inside the interaction region are longer and therefore this interaction
disturbs the correlation signal more strongly. The asymmetry in the radius
distortions is a direct proof that the appearance of the correlation does not
occur instantaneously but is a process in time.
\begin{figure}
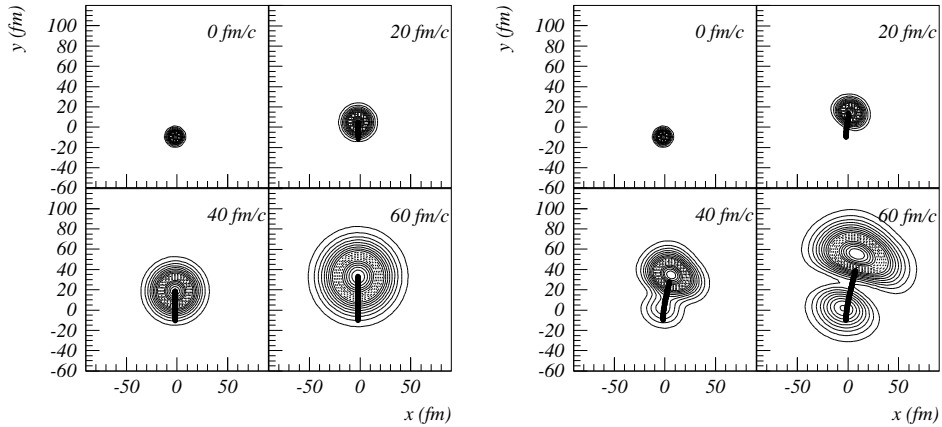

\begin{minipage}{7cm}
\epsfxsize=7.5cm
\epsffile[70 270 420 550]{cris98_fig3l.eps}
\end{minipage}
\begin{minipage}{7cm}
\epsfxsize=7.5cm
\epsffile[100 270 450 550]{cris98_fig3r.eps}
\end{minipage}
\caption{Left panel: The time development of the relative wavepacket of
unlike-charge pairs with charge $Z$ = 1. Right panel: The same for pairs with
charge $Z$ = 20.}
\end{figure}

\subsection{Coulomb Interaction between Charged Pions}
The first and the second methods were used to study the effects of their
mutual Coulomb interaction onto the two-pion correlation function. The results
were similar: The Coulomb interaction between like-charge pions alters
the correlation by such a small amount that it becomes unobservable under
normal experimental conditions~\cite{mer97a}. This conclusion is in
apparent contradiction to experimental results from ($\pi^+$,$\pi^-$)
pairs~\cite{bar97}, we will therefore focus our discussion on this type
of pairs.

The reason why the suppression of the correlation signal for very small values
of $q$ is not observed for like-charge pion pairs is connected to the zero
point energy $E_0 = \frac{3}{2} T_{qm}$ of the pion which is almost two orders
of magnitude larger than the Coulomb energy between two pions with localization
$\sigma_0$ = 1.8 fm. This is particularly easy to see in the case of two pions
with different charges where this Coulomb energy attains its maximum
value at complete overlap and is, for particles with charge number $Z$, of
size $E_{Coul} = - \frac{Z^2e^2}{\sqrt{\pi}\sigma_0}$. One may define the
parameter~\cite{mer98}
\begin{equation}
\xi = \frac{ \left| E_{Coul} \right|}{E_0} = \frac{8 e^2}{3 \sqrt{\pi} \hbar^2}
Z^2 m \sigma_0
\end{equation}
the value of which determines whether or not the Coulomb interaction becomes
discernible. For $\sigma_0$ = 1.8 fm and $Z$ = 1 one obtains $\xi = 1.4 \cdot
10^{-2}$, but $\xi = 5.6$ in case of a hypothetical charge number $Z$ = 20.
The time development of the pion wavepacket in {\bf relative} coordinates is
shown in Fig.3 for both cases, the contour lines give the shape of the
wavepacket, the dark curves the movement of its centre. It is evident that in
the first case the wavepacket movement and its deformation is not modified by
the Coulomb interaction whereas it is strongly modified in the second case
with respect to the centre as well as to the deformation. In order to observe
the Coulomb interaction the $\xi$ parameter has to be of order 1. This
condition cannot only be met by an increase of the charge number $Z$ but also
by increasing the reduced mass of the particle pair or by increasing the
localization $\sigma_0$. Finally one may also test the validity of the present
procedure by approaching the classical limit $\hbar \rightarrow 0$. These
dependences were studied in detail, as examples the Fig.4 displays in the left
panel the ($\pi^+$,$\pi^-$) correlation functions for three different values
of $\sigma_0$ and for the case $\hbar \rightarrow \hbar / 10$. In this latter
case the calculated result almost agrees with the classical expectation shown
by the continuous curve. The dependence on $\sigma_0$ is as predicted, as a
cross check, shown in the right panel, we have also performed the calculation
for $\sigma_0$ = 1.8 fm using the second method.
\begin{figure}
\epsfxsize=15cm
\epsffile[-50 390 600 650]{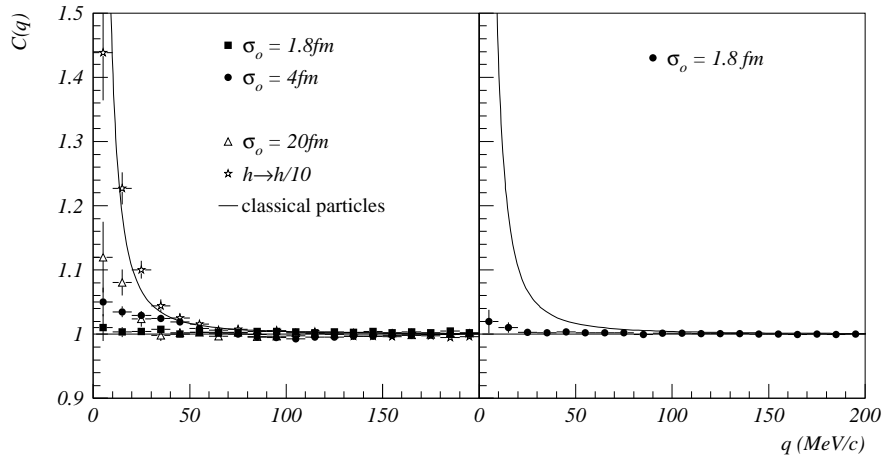}
\caption{The correlation function of ($\pi^+$,$\pi^-$) pairs with different
localizations $\sigma_0$. Also shown is the classical result (full curve) and
its limit $\hbar \rightarrow \hbar/10$. The results in the left panel are from
method 1, in the right panel from method 2 (see text).}
\end{figure}

We conclude that the Coulomb interaction only modifies the correlation function
at small values of $q$ when the $\xi$ parameter is of order one. Whereas this
is not to be expected at SIS energies in case of pion pairs, it can occur for
such pairs at much higher energies. At high energies pions are also emitted
from the decay of long-living resonances which were produced by the nucleus -
nucleus collision. Such pions have a large coherence length with the related
extended localization. The strength of the Coulomb distortion depends on the
relative abundance of pions from the interaction zone to those from the
long-living resonances, only the former will contribute to the observation of
the HBT signal. The strength of the HBT signal at $q \rightarrow 0$ also
depends on the pion localization and on the source temperature, dependences
which experimentally have not been explored up to now.
Our results illustrate the important role the pion coherence length plays in
describing the HBT interferometry with pions.

\section*{Acknowledgments}
This work was supported by the Bundesministerium f\"ur Forschung und
Technologie under contract No 06 HD 525 I and by the Gesellschaft f\"ur
Schwerionenforschung mbH under contract No HD Pel K.


\end{document}